\documentclass{PoS}

\title{HMC algorithm for two-flavour lattice QCD: Schwarz-preconditioning 
     with a one-dimensional domain decomposition}

\ShortTitle{HMC algorithm }

\author{{Martin Hasenbusch} \thanks{Address after 31.10.2007: 
 Institut f\"ur Theoretische Physik, Universit\"at Leipzig, Postfach 100 900,
 D-04009 Leipzig, Germany}\\
             Dipartimento di Fisica
     dell'Universit\`a di Pisa and I.N.F.N., 
    Largo Bruno~Pontecorvo 3, \\
  I-56127 Pisa, Italy\\
        E-mail: \email{Martin.Hasenbusch@df.unipi.it}}

\abstract{We study a variant of the Schwarz-preconditioned HMC algorithm. 
In contrast to the original proposal of L\"uscher, we apply the domain
decomposition in one lattice direction only. 
This is sufficient to reduce the condition number of the fermion matrix 
restricted to the domains compared with the full fermion matrix.
For the same linear extension of the domain, less links reside on the 
boundaries of the domains.
Therefore it becomes e.g. practical to iterate the decomposition.
We perform numerical tests for two degenerate flavours of 
Wilson fermions. The standard Wilson gauge action at $\beta=5.6$ is used. 
The performance 
of our implementation is compared with other recent studies using 
various types of preconditioning. 
}

\FullConference{The XXV International Symposium on Lattice Field Theory\\
		 July 30-4 August 2007\\
		 Regensburg, Germany}

\begin{document}
\section{Introduction}
We consider a  system with two degenerate flavours of quarks
that is defined by the partition function
\begin{equation}
\label{partition}
 Z = \int \mbox{D}[U] \exp(-S_G[U])  \;\; \mbox{det} M[U]^2 \;\; ,
\end{equation}
where 
%\begin{equation}
$
S_G[U] = - \frac{\beta}{3} \sum_{x} \sum_{\mu>\nu}
\mbox{Re} \; \mbox{Tr} \;
\left ( U_{x,\mu} U_{x+\hat \mu,\nu} U_{x+\hat \nu,\mu}^{\dag}
U_{x,\nu}^{\dag} \right ) \;\;
$
%\end{equation}
is the standard Wilson plaquette action,
$x=(x_0,x_1,x_2,x_3)$ with $x_i$ integer in the range $0 \le x_i < L_i$
are sites on a hyper-cubical lattice,
$\mu,\nu \in \{0,1,2,3\}$ are directions
on the lattice and $\hat \mu$ is a unit vector in $\mu$-direction.
The gaugefield $U_{x,\mu}$ is an element of the group $SU(3)$.
In eq.~(\ref{partition}), the
fermion degrees of freedom have been integrated out, leading to the
fermion determinant in the weight. 
The Wilson fermion matrix is given by
\begin{equation}
 M[U]_{xy} =
 1
- \kappa \sum_{\mu} \left\{
 (1-\gamma_{\mu} ) \; U_{\mu}(x) \; \delta_{x+\hat \mu,y} +
 (1+ \gamma_{\mu} ) \; U_{\mu}^{\dag}(x-\hat \mu)\; \delta_{x-\hat \mu,y}
\right\} \;\; ,
\end{equation}
where the $\gamma_{\mu}$ are the euclidian $\gamma$-matrices,
and $\kappa$ is the so called
hopping parameter, which is related with the bare mass of the fermions.

Recently there had been algorithmic progress  
\cite{MH_schwinger,PeardonLattice,karlandme,schwarz,kennedy}
in the simulation of lattice QCD at light quark masses.
In two flavour simulations, following \cite{WePe}, the determinant of 
the fermion matrix $M$ is represented as
%\begin{equation}
%\label{standardpf}
$
\mbox{det} M M^{\dag} \propto \int \mbox{D} \phi^{\dag} \int \mbox{D} \phi \;
\exp(- |M^{-1} \phi|^2) $,
%\end{equation}
where $\phi$ is the pseudo-fermion field and $S_{pf}=|M^{-1} \phi|^2$ the 
pseudo-fermion action. 
The basic idea of  
\cite{MH_schwinger,PeardonLattice,karlandme,schwarz,kennedy}
is to chose alternative representations of the 
fermion determinant while keeping the Hybrid Monte Carlo (HMC) algorithm  
unchanged otherwise. 
To this end, the fermion matrix is factorized
$M=\prod_i W_i$
such that the factors $W_i$ have a smaller condition number than the
fermion matrix $M$ itself. A pseudo-fermion field is introduced for
each of the factors
\begin{equation}
\label{factorpf}
\mbox{det} M M^{\dag} = \prod_{i=1}^{n} \mbox{det} W_i W_i^{\dag}
                      \propto 
\int \mbox{D} \phi_1^{\dag} \int \mbox{D} \phi_1 \;
\int \mbox{D} \phi_2^{\dag} \int \mbox{D} \phi_2 \;
... \int \mbox{D} \phi_n^{\dag} \int \mbox{D} \phi_n \;
\exp(- \sum_i |W_i^{-1} \phi_i|^2) \;\;. \nonumber
\end{equation}
%\begin{eqnarray}
%\label{factorpf}
%\mbox{det} M M^{\dag} &=& \prod_{i=1}^{n} \mbox{det} W_i W_i^{\dag} \\
%                      &\propto &
%\int \mbox{D} \phi_1^{\dag} \int \mbox{D} \phi_1 \;
%\int \mbox{D} \phi_2^{\dag} \int \mbox{D} \phi_2 \;
%... \int \mbox{D} \phi_n^{\dag} \int \mbox{D} \phi_n \;
%\exp(- \sum_i |W_i^{-1} \phi_i|^2)  \nonumber
%\end{eqnarray}
The effect of this splitting is two-fold: The noise of the stochastic 
representation of the fermion matrix is reduced compared with the standard
pseudo-fermion action and furthermore, the splitting of the action allows  
to compute numerically expensive parts  
less frequently, as suggested in \cite{SeWe}.

Here we discuss a variant of the Schwarz-preconditioned HMC  put forward by 
L\"uscher \cite{schwarz}. While in the other cases 
\cite{MH_schwinger,PeardonLattice,kennedy} the factors $W_i$ can be
written as a function of the fermion matrix, here a spatial decomposition 
is the basis for the factorization. 

The lattice is decomposed into blocks of the size 
$l_0 \times l_1 \times l_2 \times l_3$,
with $l_{\mu}<L_{\mu}$. %, where $L_\mu$ 
% are the linear extensions of the lattice. 
An approximation $W_1$ of $M$ is obtained by eliminating the 
hopping terms in $M$ that connect different blocks. 
L\"uscher \cite{schwarz} made the important observation that 
$\mbox{det}^2(W_1^{-1} M)$ can be estimated by using a pseudo-fermion field 
that resides on the boundaries of black blocks only (lets assume 
a red/black decomposition of the blocks.).
Furthermore in eqs.~(3.12,3.13) of  
\cite{schwarz} he shows how the force due to the pseudo-fermion action
for $\mbox{det}^2 (W_1^{-1} M)$ can be computed efficiently.  In the 
following we shall use these results without any modification; also the 
result of 
Appendix B of \cite{schwarz} is used in the following 
to reduce the dimension of the pseudo-fermion field by half.

Here we consider a block-decomposition in one dimension 
only, say the temporal direction. I.e. $l_{\mu}=L_{\mu}$ for $\mu=1,2,3$. 
The reasons to study this special case are the following:
a) the implementation becomes much simpler; mainly because there are no sites 
in a corner of block.
b) At least for the lattice spacings currently investigated, the fraction of 
links on the boundary between blocks is much less; therefore the number 
of active links, i.e. those links that take part in the 
molecular-dynamics evolution is larger.
c)
The simplification enables us to iterate the block decomposition.

Disadvantages of the one-dimensional decomposition are that it is less 
useful for a massive parallelization of the program and what might be more
important, for the same $l_0$ the condition number of $W_1$  might be 
larger than for a decomposition in all four directions. However the
experience with Schr\"odinger functional boundary conditions suggests that
still there is a substantial reduction of the condition number of 
$W_1$ compared to $M$.

%\section{The Pseudo-fermion action}
In our numerical experiments, we have iterated the decomposition 
twice. In the simulations discussed below, we have chosen $l^{(1)}_0 = L_0/2$
for the first step 
and $l^{(2)}_0=l^{(1)}_0/2=L_0/4$ for the second step of the decomposition. 
$W_i$ denotes the fermion matrix restricted to the blocks of size 
$l^{(i)}_0$. For $W_2$ we have used even-odd and 
mass-preconditioning \cite{MH_schwinger}: % for the 
%fermion matrix restricted to the blocks of size $l^{(2)}_0$:  
$W_{3,eo}$ = $W_{2,eo} + \rho$. I.e.
the pseudo-fermion action consists of four parts: 
$S_{4}$, $S_{3}$, $S_{2}$, $S_{1}$ representing the squares of the determinant
of $M W_1^{-1}$, $W_1 W_2^{-1}$, 
$W_{2,eo} W_{3,eo}^{-1}$ and $W_{3,eo}$, respectively. 
Note the counter-intuitive connection between the labels of the $S$ and the $W$.
$S_0$ is given by the gauge action.

\section{Integration with multiple time scales}
The basic steps of the integration scheme are given by
\begin{equation}
 T_U (\Delta \tau) \; : \; U \rightarrow e^{i \Delta \tau \; P }
                         \; U \;\;\;\;\; \mbox{and} \;\;\;\;\;
 T_{P,j} (\Delta \tau) \; : \;  P \rightarrow P
  - i \Delta \tau \;
                      \delta_U S_j(U) \;\;\;,
\end{equation}
where $\delta_U$ denotes a variation with respect to the gauge fields.
From these basic steps we can build elementary leap-frog  steps
\begin{equation}
T_{LF,0}(\Delta \tau_0) = T_{P,0}\left(\frac{\Delta \tau_0}{2}\right) T_U(\Delta \tau_0) 
           T_{P,0}\left(\frac{\Delta \tau_0}{2}\right) 
\end{equation}
or steps of an improved scheme (here we follow \cite{SeWe}):
\begin{equation}
\label{swimp}
 T_{SW,0}(\Delta \tau_0) = T_{P,0} \left(\lambda \Delta \tau_0 \right) \;
            T_U \left(\frac{\Delta \tau_0}{2} \right) \;
            T_{P,0} \left([1- 2 \lambda] \Delta \tau_0 \right) \;
            T_U \left(\frac{\Delta \tau_0}{2} \right) \;
            T_{P,0} \left(\lambda \Delta \tau_0 \right) 
\end{equation}
with $\lambda=1/6$.
Note that in an elementary step of this scheme, the variation of the action
with respect to the gauge-fields has to be computed twice. % Comparing the
%efficiency of this scheme and the leap-frog we have to keep this factor of two
%in mind. 
This scheme is closely related with the
second order minimum norm scheme (2MN) studied in \cite{taka05}.
The only difference is the choice $\lambda \approx 1/5$ in 
\cite{taka05}.
$S_0$ is  the part of the action with the largest forces. 
%in our case 
%it is the gauge action $S_G$. 
Elementary integration steps that include parts
$S_j$ of the action that have smaller forces are now constructed recursively as
\begin{equation}
\label{iterateLF}
T_{LF,j}(\Delta \tau_j) = 
T_{P,j}\left(\frac{\Delta \tau_j}{2}\right) \;
[T_{X,j-1}(\Delta \tau_{j-1})]^{n_{j-1}} \;
T_{P,j}\left(\frac{\Delta \tau_j}{2}\right)
\end{equation}
in the leapfrog case and 
\begin{equation}
\label{iterateSW}
T_{SW,j}(\Delta \tau_j) = 
    T_{P,j} \left(\lambda \Delta \tau_j \right) \;\;
    [T_{X,j-1} \left(\Delta \tau_{j-1} \right)]^{n_{j-1}/2} \;\;
    T_{P,j} \left([1- 2 \lambda] \Delta \tau_j \right) \;\;
    [T_{X,j-1} \left(\Delta \tau_{j-1} \right)]^{n_{j-1}/2} \;\;
    T_{P,j} \left(\lambda \Delta \tau_j \right) 
\end{equation}
in the improved case. The step sizes of the different levels are related as
$\Delta \tau_j=n_{j-1} \Delta \tau_{j-1}$. 
In both cases $X$ can be either leap frog ($LF$) or 
the improved scheme ($SW$).  This means that for different time scales, 
different integration schemes can be used. Here we have used the 
leapfrog scheme for the levels $j=2,3,4$ and the improved one for $j=0,1$. 
A full trajectory is given by $T_{LF,4}(\Delta \tau_4)^{n_4}$. 

%\subsection{step size as a function of $x_0$}
In the case of the Schwarz-preconditioning, the force due to the 
pseudo-fermion action depends quite strongly on the position of the 
gauge link with respect to the boundaries of the blocks. I.e.  here
 on $x_0$. Therefore, as discussed in \cite{schwarz}, one might 
chose a step size that depends on the position, such that the step size 
times the force is roughly constant.  As we shall see below, the force 
is the largest close to the boundaries of the blocks.  Therefore, 
we have used the following schemes: \\
(A) In the case of $L_0=24$ we have used
$s(x_0)=0.2$, $0.5$, $1$, $1$, $0.5$ and $2$ for $x_0=0$, $1$, ..., $5$
for the space-like links and
$s(x_0)=0.2$, $0.5$, $1$, $0.5$, $0.2$ and $0$ for $x_0=0$, $1$, ..., $5$
for time-like links. This scheme
is then repeated: $s(x_0+6 n)= s(x_0)$, where $n \in {1,2,3}$.
\\
(B) for $L_0=32$ is given by
$s(x_0)=0$, $0.5$, $1$, $1$, $1$, $1$, $0.5$ and $0$, 
for $x_0=0$, $1$, ..., $7$
for the space-like links and
$s(0)=0$, $0.5$, $1$, $1$, $1$, $0.5$, 
$0$ and $0$ for $x_0=0$, $1$, ..., $7$, for time-like links. 
This scheme
is then repeated: $s(x_0+8 n )= s(x_0)$, where $n \in {1,2,3}$. \\
(C) for $L_0=32$  is given by
$s(0)=0$, $0.25$, $0.5$, $1$, $1$, $1$ , $1$,
$0.25$, $0.25$, $1$, $1$, $1$, $1$,
$0.5$, $0.25$, $0$, for $x_0=0$, $1$, ..., $15$ for the spatial links and
$s(0)=0$, $0$, $0.25$, $0.5$, $1$, $1$ , $0.25$,
$0$, $0.25$, $1$, $1$, $1$, $0.5$,
$0.25$, $0$, $0$ for $x_0=0$, $1$, ..., $15$ for the time-like links. 
%The values for
For $x_0>15$: $s(x_0)=s(x_0-16)$.

Note that the blocks of the first decomposition run from $x_0=0$ up to 
$L_0/2-1$ and from $x_0=L_0/2$ up to $L_0-1$.  For the scheme (A)
the average of $s$ over all links is $0.525$.  For the schemes (B) and
(C) it is about $0.59$. 
The actual step size for a given link is $\Delta \tau$ quoted 
below times $s(x_0)$. 
In order to ensure ergodicity of the update, the configuration is shifted 
in time direction after each trajectory. 

\section{Numerical results}
We have simulated the 
Wilson gauge action at $\beta=5.6$ with
Wilson fermions using the values of the hopping parameter: 
$\kappa=0.1575$, $0.1580$  and $0.15825$. These parameters are chosen such 
that we can compare our results with 
\cite{schwarz,kennedy,urbachetal,wupper}. Following the literature, 
these bare parameters correspond roughly to a pseudo-scalar mass of 
$690$ MeV, $490$ MeV and $370$ MeV. Note that in the real world the pion 
mass is $m_{\pi} \approx 135$ MeV. The lattice spacing is about $0.8$ fm.

As solver we have used the geometric series for $S_1$, $S_2$ and $S_3$
and the BiCGstab solver with even-odd and Schwarz-preconditioning for $S_4$.
%For lack of space we do not discuss the details of the implementation 
%and the rational of this particular choice. Also, as we have learnt at the 
%conference, there might be much more efficient choices.
The basic parameters of our runs are summarized in table \ref{basictab}. 
The parameters of the algorithm have been chosen 
such that roughly the number of steps of the solver is the same for each 
part of the pseudo-fermion action.
The typical length of our runs is 2000 trajectories after equilibration 
up to about 5000 trajectories for the runs with $L=12$.  On  8 CPUs 
(Opteron 2.2 GHz) of a Cray XD1 computer one trajectory for the $32 \times 24^3$
lattice at $\kappa=0.15825$ took about 2.5 hours. Note that in our program
the Dirac operator runs with less than one Gflops per processor and the 
sub-optimal choice of solver. Our CPU time can be compared with about  $0.3$ 
hours \cite{schwarz} (from fig. 7) on 8 nodes with two 2.4 GHz
Xeon CPUs each.  Note that in this case the trajectory length is only $\tau=0.5$
and also the number of active links is about half of ours. 

\begin{table}
\caption{\sl \label{basictab}
Basic parameters of our runs. $P_{acc}$ is the acceptance rate at the end
of the trajectory. S denotes the scheme used for the $x_0$ dependence 
of the step size. $\rho$ is the parameter of the mass preconditioning. 
%as defined in eq.~(18) of \cite{karlandme}. 
%The other notations are explained in the text.
}
\begin{center}
\begin{tabular}{|l|c|l|l|l|l|l|l|l|l|l|}
\hline
$L_0$ & $L=L_1=L_2=L_3$ &S &\multicolumn{1}{|c|}{$\kappa$}&
\multicolumn{1}{|c|}{$\rho$} &  $n_4$ & $n_3$ &$n_2$&$n_1$&$n_0$ &
\multicolumn{1}{|c|}{$P_{acc}$}\\
\hline
24 & 12 &A &  0.1575 &0.15 &  6  &  1 &  2  &  2  &  4     & 0.892(2) \\
24 & 12 &A &  0.1580 &0.15 &  6  &  1 &  2  &  2  &  4     & 0.916(2) \\
32 & 16 &B &  0.1575 &0.20 &  4  &  1 &  3  &  2  &  4     & 0.704(5) \\
32 & 16 &B &  0.1580 &0.15 &  5  &  1 &  3  &  2  &  4     & 0.826(4) \\
32 & 16 &B &  0.15825&0.15 &  5  &  1 &  3  &  2  &  4     & 0.826(4) \\
32 & 24 &B &  0.15825&0.15 &  7  &  1 &  3  &  2  &  4     & 0.83(2) \\
32 & 24 &C &  0.15825&0.15 &  5  &  3 &  2  &  2  &  4     & 0.875(4) \\
\hline
\end{tabular}
\end{center}
\end{table}

\begin{figure}
\begin{center}
\scalebox{0.33}
{
\includegraphics{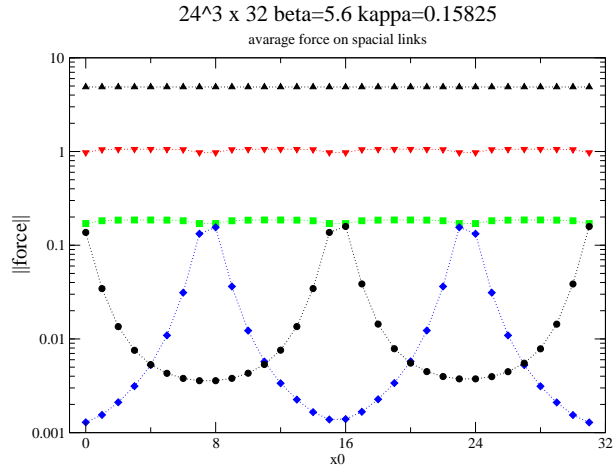}
}
\end{center}
\caption{\label{forceplot} 
We give the average force on spatial links as a function 
of $x_0$. For a discussion see the text}
\end{figure}

In fig. \ref{forceplot} we give the average forces on the spatial links
as a function of $x_0$.  The largest force is obtained for the gauge action.
The forces due to $S_3$ and $S_4$ display a strong dependence on $x_0$. 
They are largest at the boundaries between the blocks. In the case of $S_4$,
they assume their minimum in the middle of the block. In the case
of $S_3$ the minimum is located at the boundaries of the blocks
of the first decomposition.
Note that the minimum of the force due to $S_3$ is much smaller than that
of the force due to $S_4$.

The step sizes needed to obtain a sufficient acceptance rate can be compared
with results from the literature. Here we give only a small selection:
Using standard HMC, the authors of \cite{wupper} need
the step size $\Delta \tau =0.006$ for $\kappa=0.1580$ on a
$32 \times 16^3$ lattice to get $P_{acc}=0.66$. Note that in this case 
the pseudo-fermion action is computed with the fermion matrix itself 
and not with the even-odd preconditioned one.
Our most difficult case, the $32 \times 24^3$ lattice at $\kappa=0.15825$ 
we compare with  \cite{schwarz} who needs $\Delta \tau =0.05$ to reach
$P_{acc}=0.86$ and  \cite{urbachetal},
using mass preconditioning, where 
$\Delta \tau =0.1$ is needed to get $P_{acc}=0.8$. In \cite{kennedy} 
$\Delta \tau =0.25$ is used in combination with a fourth order 
minimal norm integrator. % In the case of this scheme, the force has to 
%be evaluated 5 times as often as for the leap-frog scheme that is used 
%by the other authors.

%\section{Autocorrelation times}
In order to judge the performance of an algorithm autocorrelation 
times for the quantities of interest have to be determined.  This is 
however a notoriously hard problem in HMC simulations of QCD with dynamical 
fermions. 

In fig. \ref{timehistory} 
we give the evolution of the plaquette value and the number of 
steps taken by the solver for the simulation of a $32 \times 16^3$ lattice
at $\kappa=0.1575$. The run started from a configuration equilibrated 
by a different version of HMC algorithm.  The plots give no indication 
for autocorrelation times that are comparable with the length of 
the run itself. We get $\tau_P=8(2)$ and $\tau_{solv}=16(5)$ as integrated 
autocorrelation times of the plaquette and the number of steps of the solver. 
The time unit is given by a trajectory. 
These numbers can be compared with $\tau_P=7(4)$ and $\tau_{solv}=33(4)$
for a standard HMC simulation \cite{wupper} % with acceptance rate of 
% $P_{acc}=0.73$
and  $\tau_P=68(25)$ and $\tau_{solv}=168(42)$ 
for a Schwarz preconditioned HMC simulation \cite{schwarz}. 
% and acceptance rate 
%$P_{acc}=0.81$.  
Note that in \cite{schwarz} the trajectory length is $\tau=0.5$
and only about $37 \%$ of the links are active. This might trivially 
explain a factor of about 4 compared with our simulation.  One also should 
note that the authors of \cite{Harvey} find that even larger trajectory
lengths such as $\tau=2$ are advisable to obtain optimal performance.

\begin{figure}
\scalebox{0.3}
{
\includegraphics{ene16k1575.eps}
}
\phantom{x}
%\end{figure}
%\begin{figure}
\scalebox{0.3}
{
\includegraphics{solv16k1575.eps}
}
\end{figure}
\begin{figure}
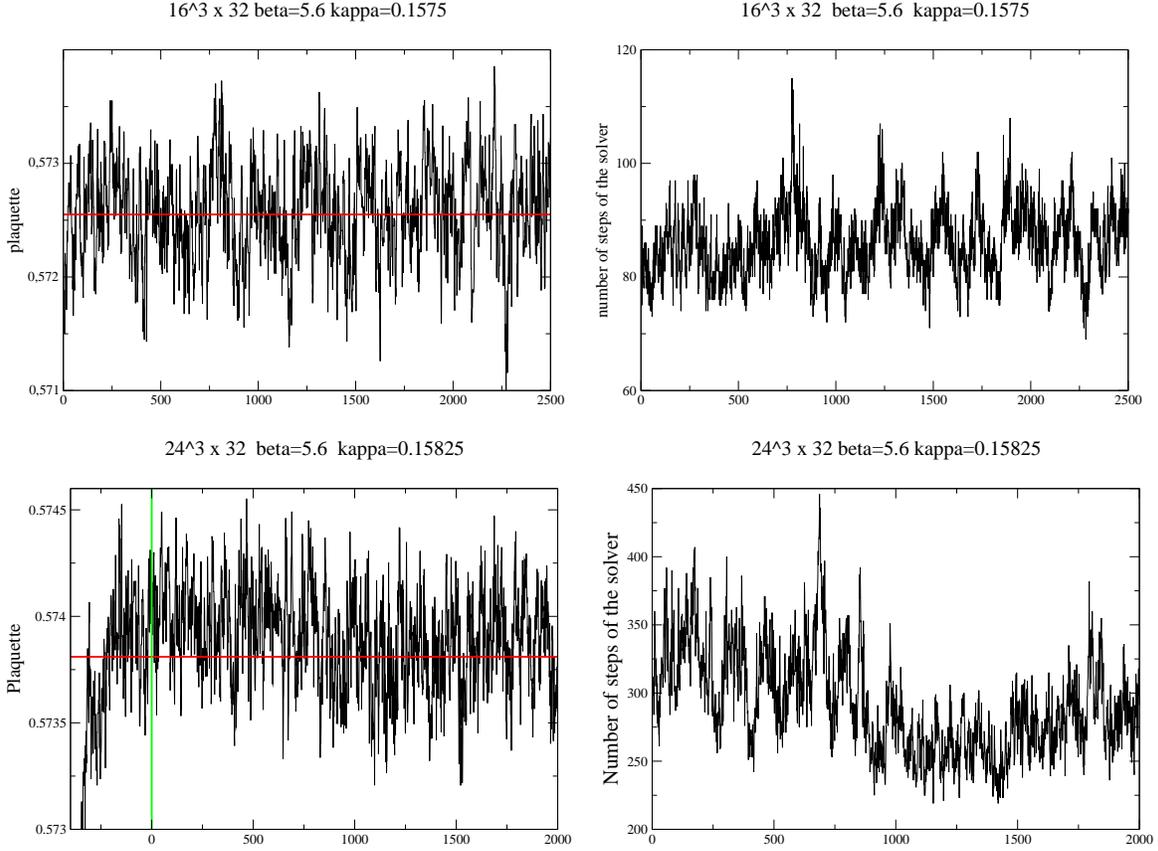

\scalebox{0.3}
{
\includegraphics{plaq24k15825.eps}
}
\phantom{x}
%\end{figure}
%\begin{figure}
\scalebox{0.3}
{
\includegraphics{solver24k15825.eps}
}
\caption{\label{timehistory} 
History of the plaquette average and the number of steps of the 
solver. The red line gives the average of the plaquette obtained in 
\cite{wupper} in the case of $L=16$ and $\kappa=0.1575$  and 
\cite{schwarz,Debbio:2005qa} in the case of $L=24$ and $\kappa=0.15825$.
}
\end{figure}

In the case of $L=24$ and $\kappa=0.15825$ we do not quote values for 
autocorrelation times. The time histories of the average plaquette and the 
number of solver steps suggests that there are correlations that are 
comparable with the length of our run or even larger. 
Note that the authors of \cite{schwarz,kennedy,urbachetal,Debbio:2005qa} 
do not see such 
problems and quote rather small values of the autocorrelation times.
Taking into account the length of the trajectory and the fraction of 
active links, our run is of similar length as that of 
\cite{Debbio:2005qa}. 
%Given the fact that our implementation seems 
%to  have smaller autocorrelation times for
%$L=16$ and $\kappa=0.1575$
 One should take into account the possibility that
\cite{schwarz,kennedy,urbachetal,Debbio:2005qa} do not see these 
large autocorrelations since their runs are too short.

\section{Conclusions and outlook}
Using preconditioned pseudo-fermion actions 
\cite{MH_schwinger,PeardonLattice,karlandme,schwarz,kennedy} the problem
that 
the step size needed to obtain a reasonable acceptance rate
decreases with decreasing fermion mass seems to be
overcome. The performance of the different proposals seems to be quite 
similar. Still the dependence of autocorrelation
times related to small eigenvalues of the fermion matrix on the choice 
of the pseudo-fermion action is not well understood.
To this end, it might be useful to monitor e.g. the topological charge. Likely 
also much longer runs then those presented here and 
in \cite{kennedy,urbachetal,Debbio:2005qa} are 
needed to this end. 
A disadvantage of the Schwarz-preconditioning is that it is quite 
hard to implement fermion actions  that are more complicated than 
clover-improved Wilson fermions.
Since in the case of Schwarz-preconditioning the pseudo-fermions 
reside on boundaries only, it is possible that the performance of the HMC 
scales differently (hopefully better) with the lattice spacing than for 
the other types of preconditioning.

\end{document}